\documentstyle[prb,aps,multicol,epsf]{revtex}

\begin{document}

\newcommand{\npi}{\hspace{-0.5cm}} 
\newcommand{\fg}[1]{Fig.~\ref{#1}}

\title{Semiconductor quantum dot - a quantum light source of multicolor photons with tunable statistics}
\author{D.V. Regelman, U. Mizrahi, D. Gershoni, E. Ehrenfreund}
\address{Solid State Institute and Physics Department,
Technion- Israel Institute of Technology, Haifa 32000, Israel}
\author{ W.V. Schoenfeld, and P.M. Petroff }
\address{Materials Department, University of California, Santa Barbara, CA 93106, USA}
\maketitle

\begin{abstract}
We investigate the intensity correlation properties of single
photons emitted from an optically excited single semiconductor
quantum dot. The second order temporal coherence function of the
photons emitted at various wavelengths is measured as a function
of the excitation power. We show experimentally and
theoretically, for the first time, that a quantum dot is not only
a source of correlated non-classical monochromatic photons but is
also a source of correlated non-classical \emph{multicolor}
photons with tunable correlation properties. We found that the
emitted photon statistics can be varied by the excitation rate
from a sub-Poissonian one, where the photons are temporally
antibunched, to super-Poissonian, where they are temporally
bunched.
\end{abstract}

PACS numbers: 78.67.Hc, 42.50.Dv, 78.55.Cr, 85.30.Vw


Semiconductor quantum dots have been extensively investigated recently
as a potential, technology-compatible quantum light emitters.
\cite{Imamoglu_NAT,Yamamoto,Imamoglu_SCI}
Such emitters are important for possible future quantum computing
\cite{QuantumComp} and cryptography.\cite{crypto}
It has been recently demonstrated that under continuous wave (cw) excitation, a
single quantum dot emits antibunched photons obeying a sub-Poissonian
statistics \cite{Imamoglu_PRB}, while under optical pulse excitation, they
emit single photon per each excitation
pulse.\cite{Yamamoto,Imamoglu_SCI,Erez_PRB2}.
Similar effects were previously observed also in optical studies of
the fluorescence from single atoms and molecules.\cite{Moerner,Orrit}

In this work we report on measurements of temporal correlations
among multi-color photons
emitted from cw optically excited single semiconductor quantum
dots (SCQD). We show for the first time, that there is a tunable
intensity correlation among photons emitted at the same and at
different wavelengths due to the recombination of excitons from
different collaborative quantum states. We show that the temporal
correlations among the photons emitted by the SCQD change
dramatically with the excitation power. While strong
characteristic antibunching correlations are observed for low
power excitations, these correlations disappear with the increase
in the excitation power and gradually transform into bunching
correlations for yet higher excitation power. These observations
demonstrate that a multiply populated quantum light source may
emit bunched photons, obeying super-Poisson statistics.

We quantitatively account for the experimentally measured
distribution of the time interval, $\tau$, between consecutively
emitted photons. Specifically, we explain the changes in the
distribution under variable excitation powers, both for photons
originating from the same spectral line, as well as for photons
from two different spectral lines. We do that by analytically
solving a set of coupled rate equations
\cite{Erez_PRB2,Erez_PRB1} describing the conditional probability
that a photon is emitted from a collective state of $j$ confined
electron-hole (e-h) pairs (i.e. the $j^{th}$ multiexciton)
following a photon emission event from a collective state of
$i>j$ e-h pairs.

The SCQD sample was grown by molecular beam epitaxy of a strained
epitaxial layer of InAs on (100) oriented GaAs substrate. Small
islands of In(Ga)As connected by a very thin wetting layer are
thus formed in the Stranski-Krastanov growth mode. The vertical
and lateral dimensions of the InAs SCQDs were adjusted during
growth by the partially covered island growth
technique.\cite{Garcia_APL} The sample was not rotated during the
growth of the strained layer, therefore a gradient in the QDs
density was formed and low density areas, in which the average
distance between neighboring QDs is larger than our optical
spatial resolution, could easily be found on the sample surface.

We use a diffraction limited low temperature confocal optical
microscope\cite{Erez_PRL} for the photoluminescence (PL) studies
of the single SCQDs. In order to measure the temporal correlation
between emitted photon pairs we constructed a wavelength
selective Hanbury-Brown and Twiss (HBT) \cite{HBT} setup (see
inset, Fig. 2b) including a beam splitter (BS) which divides the
PL emission from the SCQD into two equal beams. Each beam is then
dispersed by a 0.22m monochromator (MC1 and MC2) and focused into
a thermo-electrically cooled, avalanche silicon photodiode (D1
and D2). Each time a photon is detected in one of the detectors,
an electronic pulse is produced and transmitted to a time to
analog converter. The temporal difference between a pair of
pulses is translated into an output pulse of proportional voltage
while a multichannel analyzer sorts the pulses and produces
histograms of the number of counts as a function of the time
between pairs of detected photons. The temporal resolution of our
system, as evaluated by its response to a picosecond laser pulse
is $\simeq$250 picoseconds. By tuning the monochromators we can
correlate two photons emitted not only from the same spectral
line (auto-correlations), but also from different spectral lines
(cross-correlations).

We used a Ti:Sa laser at 1.6 eV, in order to non-resonantly excite
carriers in the sample. The SCQDs are thus
populated by diffusion of the photogenerated carriers into them. We locate
an optically excited SCQD by scanning the sample surface while monitoring
the resulted PL spectra using a 0.22m monochromator followed by a
liquid nitrogen cooled charge coupled device (CCD) array detector.
Once a typical SCQD emission spectrum is observed the scan is
terminated and the objective position is optimized above the SCQD.
The PL emission spectra from a single quantum dot for increasing
cw excitation powers are shown in Fig. 1. The measured spectra
strongly depend on the excitation power because of the
shell-filling effect and the Coulomb interactions between the
carriers\cite{Erez_PRL,Landin_SCI,Hawrylak_NAT}. For the lowest
excitation power there is a finite probability for one e-h pair
(exciton) to occupy the dot. The radiative recombination of this
pair gives rise to a sharp line (denoted as $X^{0}$). As the
excitation power is increased, the probability to find few e-h
pairs in the dot increases significantly. Since only two carriers
with opposite spins can occupy each non-degenerate level, a
second level is occupied when there are three or more excitons in
the SCQD. The radiative recombination of carriers occupying the
first and second energy levels, gives rise to the emission of
photons in two groups, denoted S and P, respectively, in analogy
with atomic shells. The Coulomb exchange interaction reduces the
effective bandgap of an optically excited SCQD (in a similar way
to the well known bulk phenomenon of bandgap renormalization) and
gives rise to red shifted PL lines \cite{Erez_PRL,Hawrylak_NAT}
(denoted by $nX^{S}$ ), when $n>1$ spectator e-h pairs are present
in the SCQD during the recombination.

At yet higher excitation powers, the probability to find higher
number of e-h pairs within the QD increases and the probability to
find the QD with a small number of e-h pairs decreases. As a
result, the intensity of any emission line under cw excitation
increases, reaches a maximum and then decreases as the excitation
power is further increased.

By using our wavelength selective HBT setup, we measure the intensity
correlation function between photons emitted at various wavelengths, due to the
recombination of e-h pairs in the presence of different number of spectator
other pairs. The measured correlation function is expressed as,
\begin{equation}
g_{ij}^{(2)}(\tau )=\frac{\langle I_{i}(t)I_{j}(t+\tau )\rangle }{%
<I_{i}(t)><I_{j}(t)>}~,  \label{a}
\end{equation}
where $I_{k}(t)$ is the emission intensity at a wavelength
corresponding to the recombination of a $k^{th}$  multiexciton at
time $t$. The function $g_{ij}^{(2)}(\tau )$ represents
therefore  the conditional probability that a photon from
recombination events which involves $j$ e-h pairs will be emitted
at time $\tau$ after such an emission which involve $i$ pairs has
previously occurred. Obviously, when the same spectral line is
monitored on both channels $(i=j)$, Eq. 1 simply turns into the
second order temporal coherence function. We note that completely
uncorrelated photons have $ g^{(2)}(\tau)=1$ (Poisson
statistics), photons with positive correlation (bunched photons)
have $ g^{(2)}(\tau )>1$ (super-Poisson statistics), and photons
with negative correlation (antibunched photons) have
$g^{(2)}(\tau )<1$ (sub-Poisson statistics). It is very well
established \cite{Loudon} that chaotic and thermal light sources
are characterized by $g^{(2)}(0) > 1$ and $g^{(2)}(\tau) <
g^{(2)}(0)$, while quantum light sources are characterized by
$g^{(2)}(0) < 1$ and $g^{(2)}(\tau) > g^{(2)}(0)$.

The measured intensity auto-correlation function for the $X^{0}$
spectral line ($g_{11}^{(2)}(\tau )$) for an increasing
excitation powers is presented in Fig. 2a. The figure
demonstrate that for low excitation power the probability of
simultaneous detection of two photons is zero (the photons are
antibunched\cite{Imamoglu_PRB}). This can be readily understood
intuitively, since the recombination of a single e-h pair, which
results in the detection of a photon from the $X^{0}$ line,
empties the SCQD. The probability to detect another photon
immediately after the detection of the first one is then zero,
since the time it takes for the SCQD to repopulate and emit
another photon depends on the excitation power and the e-h pair
lifetime. Thus, as can be seen in Fig. 2a, this population
regeneration time (the width of the antibunching notch) decreases
as we increase the excitation intensity. For further increase in
the excitation power, the population regeneration time continues
to decrease, as a result the measured $g_{11}^{(2)}(0)$, which is
limited by the temporal resolution of our setup, ceases to vanish.
With yet further increase of the excitation power, as the $X^{0}$
line emission intensity decreases (see Fig. 1a) the emitted
photons appear to be \emph{bunched}. This novel observation can
intuitively be understood as follows. The $X^{0}$ line
auto-correlation function $g_{11}^{(2)}(\tau )$ reflects the
probability to find the SCQD occupied with a single e-h pair at
time $\tau $ after the emission of previous $X^{0}$ photon, which
actually left the SCQD empty. At high excitation powers the
average number of e-h pairs occupying the dot in steady state is
large. Therefore, the average probability to find the SCQD
occupied with only a single pair is small\cite{Erez_PRB2}.
However, a short time after the SCQD was emptied from pairs (as
evident by the detection of the first $X^0$ photon), this
probability increases, becoming thus larger than the average one.

From the above discussion, it naturally follows that there are
also strong temporal correlations between the emission of photons
due to recombination of a given collective state of e-h pairs and
that from any other collective state. These correlations depend
on the dynamics of the SCQD population regeneration. In Fig. 2b
we present the measured cross-correlation between the emitted
photons from the $nX^{S}$ line and the photons from the $X^{0}$
line for various excitation powers. While the auto-correlation
function (Fig. 2a) is symmetric, the cross-correlation function
is very asymmetric. At low excitation power this asymmetry
reflects the fact that the elapsed time between the emission of
an $X^0$ photon (after which the SCQD is empty) and the emission
due to recombination of one pair out of the collective $n$ pair
state, is larger than the reverse order of events. At higher
excitation power, the cross-correlation function even changes
sign. This means that the probability that a single pair
recombines in the SCQD short time after the recombination of one
pair out of $n$ ($n\approx 3$) has occurred, is larger than the
product of the average probabilities, while the probability for
the opposite order of events is smaller than the product of the
average probabilities. This peculiar behavior can also be
understood intuitively. The average number of e-h pairs in the
SCQD at this excitation power is larger than $n\approx 3$.
Therefore, the probability that a single pair recombines within
the SCQD short time after the detection of recombination from the
$n$ pair collective state is larger compared with that
probability at long times (the product of average probabilities).
While after the SCQD has been found empty, as evident by the
detection of the $X^0$ photon, it takes time for the SCQD to
repopulate. During that time no emission from the $nX^S$ line, is
likely.

We now present a quantitative analysis for the above
observations. For that purpose, we model the correlation
properties of photons emitted from a single SCQD by the
multiexcitonic coupled rate equations used
earlier\cite{Erez_PRB2,Erez_PRB1}.
We denote by $n_i$ the probability to find the SCQD occupied with
the $i^{th}$ multiexciton and by a vector
$\overrightarrow{n}(t) = (n_0(t), n_1(t),...n_N(t))$
a given multiexcitonic state of the SCQD at time t.
Here $n_0(t) $ represents the probability to find the SCQD empty
and N is the highest multiexciton with non-vanishing probability.
At non resonant cw excitation the set of coupled rate equations
which describes the temporal evolution of the SCQD
population\cite{Erez_PRB1} can be expressed in matrix form as follows.
\begin{equation}
\frac{d\overrightarrow{n}(t)}{dt}=\Psi \overrightarrow{n}(t)~,  \label{3}
\end{equation}
where the time indenpendent matrix $\Psi$ is given by,
\begin{equation}
\Psi=
\left(
\begin{array}{ccccc}
-(G+{1\over\tau_{0}}) & 1\over\tau _{1} &  &  &  \\
G & -(G+{1\over\tau _{1}}) & {1\over\tau _{2}} &  &  \\
& G & -(G+{1\over\tau _{2}}) & 1\over\tau _{3} &  \\
&  &  & \cdot \cdot \cdot &  \\
&  &  & G & -{1\over\tau _{N}}
\end{array}
\right)
\end{equation}

Here $\tau_{i}$ are the $i^{th}$ multiexciton decay times ($\tau _{0}=\infty $)
and $G$ is the cw e-h pairs photogeneration rate.
The steady state solution to
Eq. 2  $\Psi \overrightarrow{n}(t)=0$ is given by\cite{Erez_PRB2},

\begin{eqnarray}
n^{ss}_{i}=n^{ss}_{0}G^{i}\prod_{j=1}^{i}\tau _{j}~ \nonumber ,\\
n^{ss}_{0}=(1+\tau _{1}G+\tau _{1}\tau
_{2}G^{2}+...+G^{N}\prod_{j=1}^{N}\tau _{j})^{-1}~.  \label{5}
\end{eqnarray}

A general solution for Eq. 2 is given by\cite{BoyceDiprima},
\begin{equation}
\overrightarrow{n}(t)=\sum_{k=1}^{N}c_{k}\overrightarrow{\xi _{k}}e^{r_{k}t}~,
\label{10}
\end{equation}
where $\overrightarrow{\xi _{k}}$ is the eigenvector
corresponding to the eigenvalue $r_{k}$ of the matrix $\Psi $,
and the constant coefficients $c_{k}$ are determined by the
initial conditions of the problem. Note that because $\det (\Psi
)=0$, $r=0$ is always one of its eigenvalues. Thus for $t
\rightarrow \infty$ the solution $\overrightarrow{n}(t)$ is given
by the steady state solutions Eq. 3. This means that
$<\overrightarrow{n}(t)>=\overrightarrow{n}$.

Assuming for simplicity that the decay of different multiexcitons
result in different spectral lines\cite{Erez_PRB1}, we have:

$I_{i}(t)={n_{i}(t)/\tau_i}$. The function
$I_{j}(t+\tau )$ in Eq. 1 can now be expressed in terms of
$n^{i}_j(\tau)$ which is calculated analytically for each $\tau$
by solving Eq. 2 with the appropriate initial conditions:
$n_{k}(t)=\delta _{(i-1),k}$
. These initial conditions reflect the important fact that a radiative decay
of the $i^{th}$ multiexciton at some time $t$ results in a probability 1 to
find the SCQD populated by the $(i-1)^{th}$ multiexciton state at that time.
Since $I_j(t+\tau)$ is therefore independent of $t$, the correlation
function $g_{ij}^{(2)}(\tau)$ can be now expressed as,
\begin{equation}
g_{ij}^{(2)}(\tau >0)=n^{i}_{j}(\tau )/n^{ss}_{j},
g_{ij}^{(2)}(\tau <0)=g_{ji}^{(2)}(-\tau )~. \label{11}
\end{equation}
For low enough excitation power, when the probability of the dot
to be occupied by more than one exciton is negligibly small,
we find for the auto-correlation function a relatively simple
formula: $g^{(2)}(\tau)=1-exp(-\tau(1/\tau_1+G))$
provided that $N=1$. It is important to note here that it is
incorrect to use this last formula for higher excitation levels,
where $N>1$.\cite{Imamoglu_PRB} In order to calculate the
intensity correlation functions between the various SCQD emission
lines, one has to know the multiexcitonic recombination rates
$\tau_i$. We directly measured $\tau_{1}$ ($\tau_{1}$$\approx$1
nsec) and used a model to estimate the decay rates of higher
multiexcitons.\cite{Erez_PRB2,Erez_PRB1,Erez_SSC} We note, that
our solution to Eq. 2 is quite robust and that its general
behavior is not strongly dependent on the specific $\tau_i$.

The calculated auto- and cross-correlation functions convoluted
with the system impulse response function are presented by solid
lines overlaid on the experimental data in Fig. 2. General good
agreement with the experimentally measured data is obtained, by
slightly adjusting $G$ within the experimental uncertainties, to
best fit the experimental data. As can be seen our calculated
correlation functions describe quantitatively the following
phenomena:\\
(a) The crossover from antibunching to bunching for the
auto-correlations of monochromatic photons as the excitation power
increases.\\
(b) The asymmetric antibunching-bunching behavior of the
cross-correlation function between different wavelength photons and its evolution with excitation power.\\

In conclusion, we have demonstrated that the statistical
properties of photons emitted from a cw excited single
semiconductor quantum dot can be variably controlled by the
external excitation intensity. For same color photons the quantum
dot acts as a quantum light source, although the emitted photons
can be either bunched (i.e. super-Poissonian statistics) or
antibunched (i.e. sub-Poissonian statistics) in time. We have
also measured for the first time temporal correlations between
photons of different colors, and developed a semiclassical model
which quantitatively accounts for our observations.

\hspace{0.5cm} {\it Acknowledgments} --
The work was supported by the Israel Science Foundation
(335/99)
and by the Israel-US Binational Science Foundation
(453/97).

\npi {\bf Figure Captions}

\newcounter{fg}
\refstepcounter{fg}
\label{fig1}
\refstepcounter{fg}
\label{fig2}

\begin{list}
{\fg{\arabic}}{\usecounter{fg}}

\item[\fg{fig1}]
PL spectra from a cw excited SCQD for various excitation powers.

\item[\fg{fig2}]
a)Measured (thick solid lines) and calculated (thin solid lines, see text)
temporal intensity auto-correlation function of the $X^0$ spectral
line for various excitation powers. b)Measured and calculated
temporal intensity cross-correlation function of the $X^0$ and
$nX^{S}$ spectral lines for various excitation powers. Inset:
schematic description of the modified HBT setup.

\end{list}

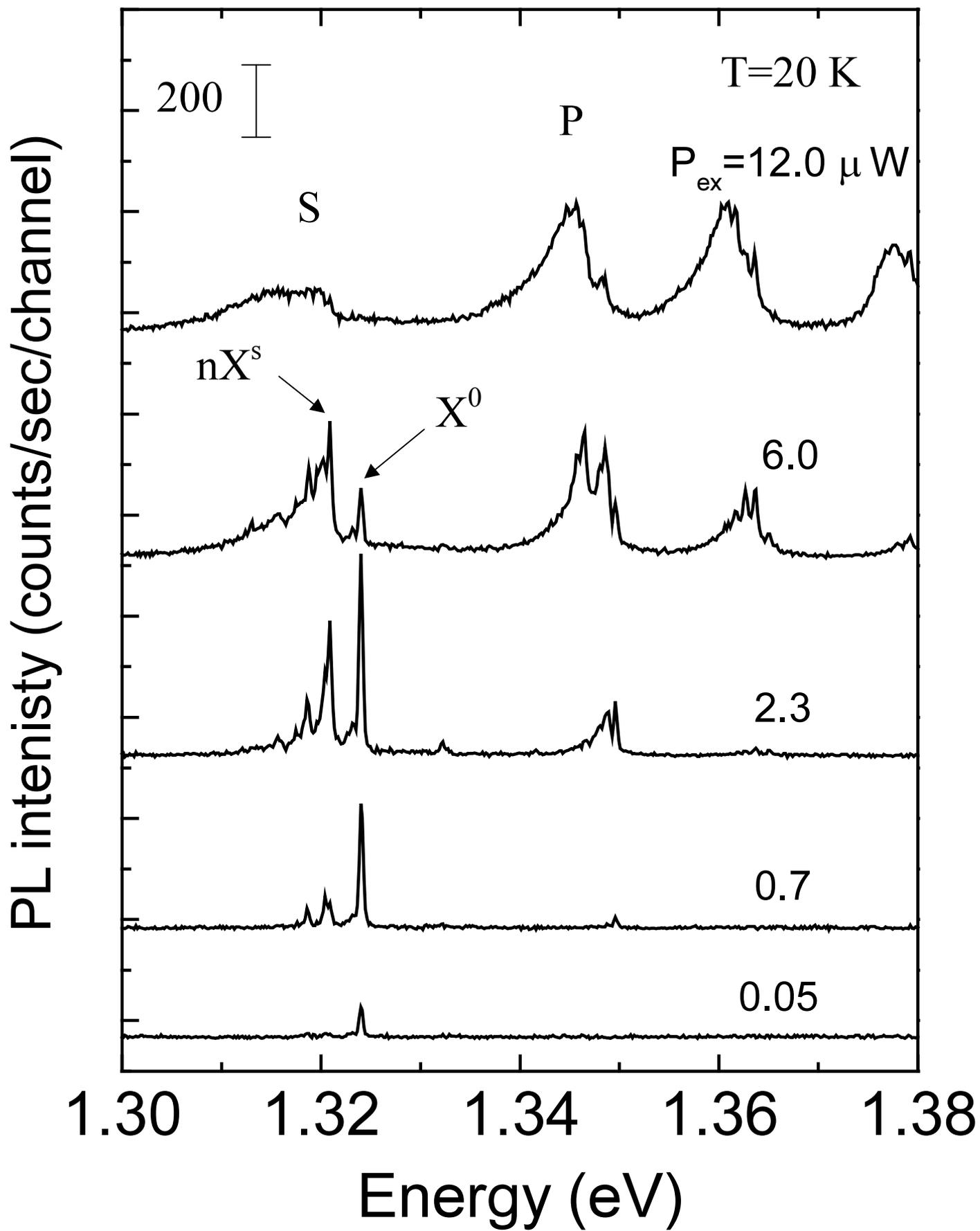
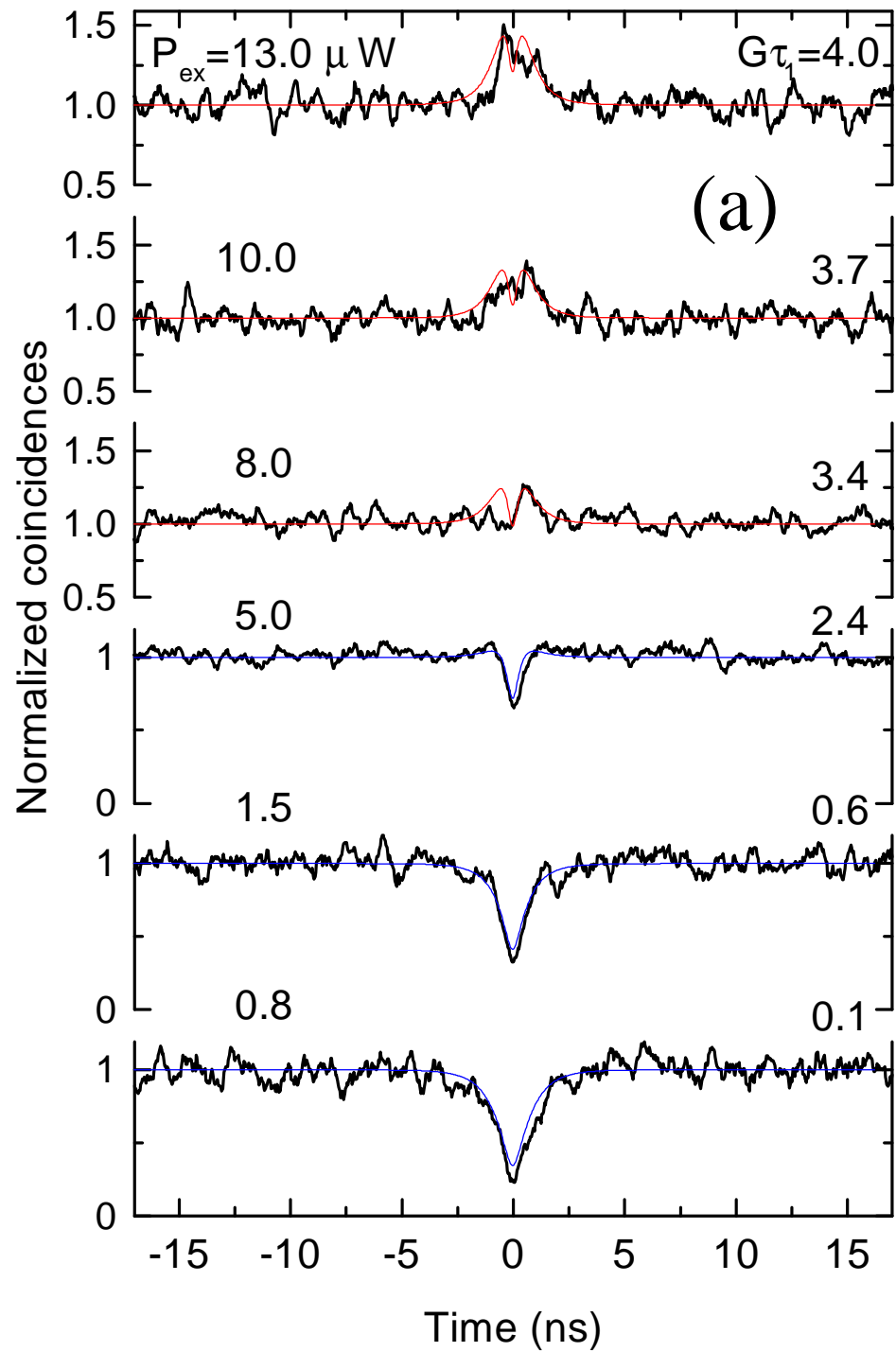
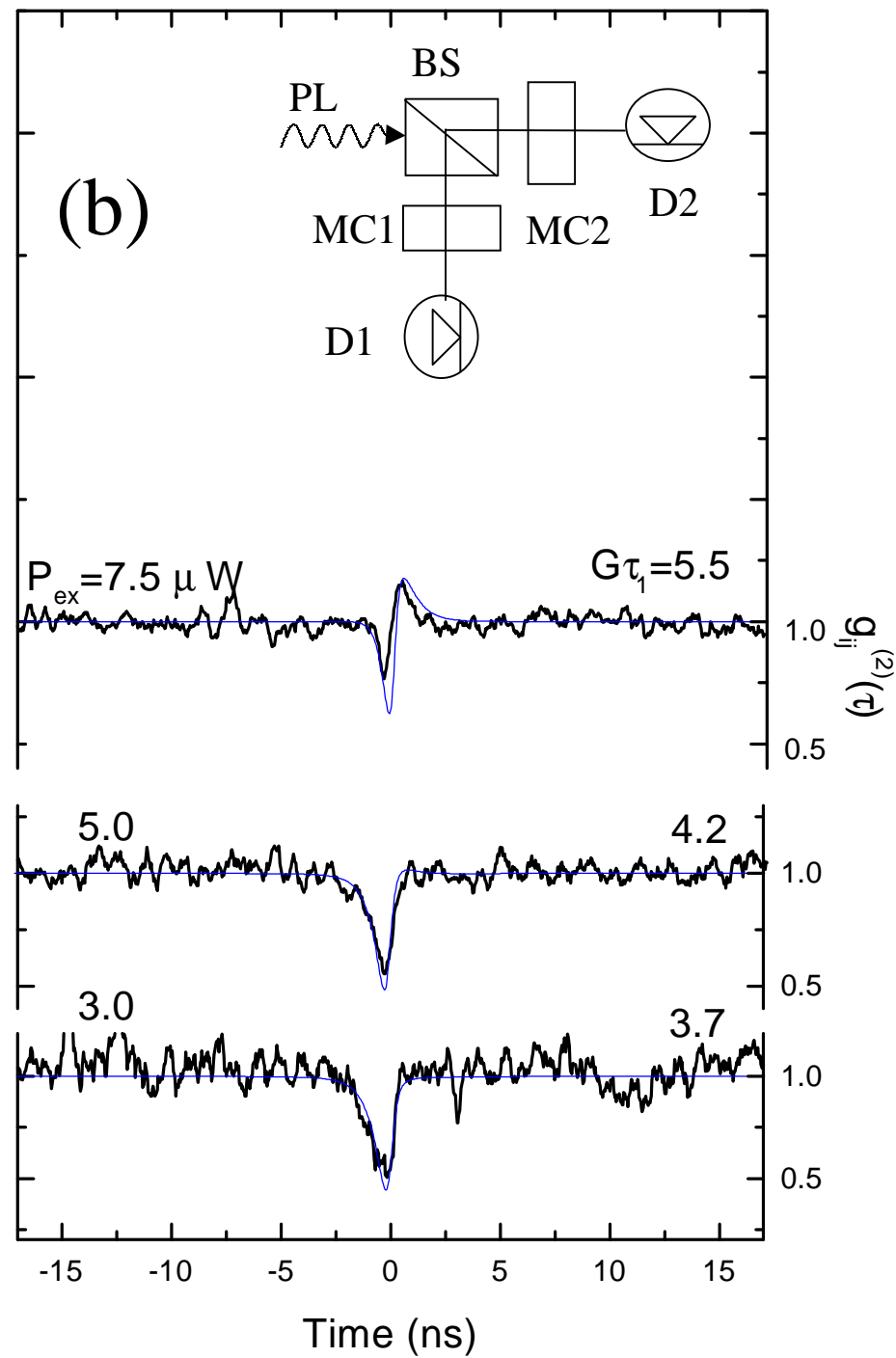
\end{document}